# ARE SANCTIONS FOR LOSERS? A NETWORK STUDY OF TRADE SANCTIONS

# ДА ЛИ СУ САНКЦИЈЕ ЗА ГУБИТНИКЕ? МРЕЖНИ ПРИСТУП ПРОУЧАВАЊУ ТРГОВИНСКИХ САНКЦИЈА

**Fabio Ashtar Telarico**
University of Ljubljana – Faculty of Social Sciences, Slovenia
Fabio-Ashtar.Telarico@fdv.uni-lj.si
ORCID: 0000-0002-8740-7078

***Abstract:*** *Studies built on dependency and world-system theory using network approaches have showed that international trade is structured into clusters of 'core' and 'peripheral' countries performing distinct functions. However, few have used these methods to investigate how sanctions affect the position of the countries involved in the capitalist world-economy. Yet, this topic has acquired pressing relevance due to the emergence of economic warfare as a key geopolitical weapon since the 1950s. And even more so in light of the preeminent role that sanctions have played in the US and their allies' response to the Russian-Ukrainian war. Applying several clustering techniques designed for complex and temporal networks, this paper shows that a shift in the pattern of commerce away from sanctioning countries and towards neutral/friendly ones. Additionally, there are suggestions that these shifts may lead to the creation of an alternative 'core' that interacts with the world-economy's periphery bypassing traditional 'core' countries such as EU member States and the US.*

***Апстракт:*** *Студије засноване на теорији зависности и светског система користећи мрежне приступе показале су да је међународна трговина структурисана у кластере 'језгра' и 'периферних' земаља које обављају различите функције. Међутим мало њих је користило ове методе да би истражило како санкције утичу на положај земаља укључених у капиталистичку светску економију. Ипак ова тема је постала хитна због појаве економског сукоба као кључног геополитичког оружја 1950-их и још више у светлу превасходне улоге коју су санкције одиграле у одговору САД и њихових савезника на руско-украјински рат. Примењујући неколико техника груписања дизајнираних за сложене и временске мреже, овај рад показује помак у обрасцу трговине од земаља које санкционишу ка неутралним/пријатељским земљама. Поред тога постоје сугестије да ове промене могу довести до стварања алтернативног „језгра" које је у интеракцији са периферијом светске економије заобилазећи традиционалне „језгро" земље као што су државе чланице ЕУ и САД.*

***Кључне ријечи:*** *Међународна трговина, Динаминче мреже, Блок моделовање, Русија, Иран, теорија светског система, санкције*

## 1. INTRODUCTION: SANCTIONS AND ECONOMIC WARFARE IN A WORLD-SYSTEM PERSPECTIVE

This paper's research puzzle stems from the contrast between two realities. On the one hand, the impedingly uncertainty regarding economic sanctions' effectiveness and a forming awareness of their inherent complexity (Morgan, Syropoulos, and Yotov 2023, 21ff). On the other, the dramatic increase in the speed at which states impose new sanctions since 2001 (cf. Felbermayr et al. 2020). In practice, notwithstanding over forty years of research, 'no consensus has yet emerged on the sign and significance of the impact of the key variables that theoretically determine the success of economic sanctions.' (Bergeijk et al. 2019, 79) Moreover, 'there is substantial uncertainty about whether such sanctions affect economic outcomes' (Felbermayr et al. 2020, 35). Thus, studies and commentaries on economic sanctions have failed to determine how structural factors alter sanctions' effects.

Arguably, the incomplete understanding of sanctions stems from the underlying theoretical framework and the related methodological constraints. Hence, this paper contributes to the existing literature on international economic sanctions by addressing the issue from the perspective of world-system analysis. Arguably, dependency/world-system theory is uniquely apt to the investigation of political-economic ties amongst sovereign political units. Moreover, this approach's relational ontology combines perfectly with network-analysis methodology. Indeed, networks have already been used to study the world-system in the last quarter of the $20^{th}$ century, but not in relation to sanctions specifically. Furthermore, both fields have advanced significantly since then making previously intractable statistical analysis possible and providing words for previously unknown geoeconomic phenomena. So, the paper applies the statistical method called *stochastic blockmodeling* (SBM) to dynamic networks of world trade.

For the sake of brevity, the discussion emphasises findings relating to sanctions against the Russian Federation (RF). But the cases of Iran and, despite their shorter duration, Venezuela are also instructive and mentioned in passing. In this way, the paper answer standing interrogative regarding how international sanctions against the RF have affected the international economy's structure since 2014, if at all.

Ultimately, the results show that sanctioned countries' trade does not dry up, especially for commodity producers such as the abovementioned countries. Instead, it shifts from sanctioning countries towards friendlier/neutral ones. Importantly, this change seems to induce no appreciable deterioration of the target countries' position in the world economy.

Rather, the cumulative effect of these shifts may be leading to the creation of an alternative 'core' that bypasses traditional 'core' identifiable with the 'West' (i.e., the US and its allies).

The paper is organised as follows. The second section provides a brief primer on the international sanctions against the RF. The third section provides basic theoretical framing within the context of world-system theory and formulates a first hypothesis.

The four section explores the existing literature on sanctions concisely, deriving four more hypothesis that operationalise the research question. The fifth section introduces the data and the methods employed in this analysis. The sixth section explores the finding in light of the five hypotheses identified previously. The last section summarises the most important findings and highlights how future researches can complement this paper's limitations,

## 2. BACKGROUND

The use of sanctions became commonplace in foreign policy during the so-called 'War on Terror', when 'financial warfare' took its current shape (Zarate 2013, 44). But the most economically relevant, politically significant, and scientifically poignant slate of sanction only emerged after Crimea became part of the RF in 2014. And those were only the first skirmishes in an economic war that sent the number of new sanctions skyrocketing in 2021/2 (Morgan, Syropoulos, and Yotov 2023, 11).

Overall, the US and its allies (the 'West') have struck the RF with three batches of sanctions (cf. Korhonen, Simola, and Solanko 2018 for more details). First, in July 2014, the senders outlawed the trade of military and dual-use goods, cut Russian entities off the European Bank for Reconstruction and Development, and banned long-term loans to state-owned banks and financial institutions.

Then, in 2018, the sanctions struck the energy sector (Rosneft, Transneft, and Gazpromneft) and the entire military-industrial complex. Third, after Russia's invasion of Ukraine, the US and its allies imposed unprecedented new sanctions on the RF (see Notermans 2022).[1]

## 3. THEORETICAL FRAMEWORK

In a nutshell, world-system theory argues that, at any given point in time, capitalism structures itself as a world-system (WS). Schematically (Wallerstein 1976, 229–33), a WS is a set of cross-border economic relations articulated around a value chain that involves several, distinct human groups and which is substantially independent from the external environment.

---

[1] So much so, that the senders had to evoke a rather obscure provision of the *General Agreement on Trade and Tariffs* to justify restrictions against another WTO member (Chachko and Heath 2022).

Thus, it is an economic system that constitutes a *world* in and of itself, but it is not necessarily *global*. Remarkably, the political-economic relations constitutive of the world-system are chiefly interactions amongst sovereign political entities. Finally, each actor plays a 'role' in these international connections that determines its ability to appropriate newly generated wealth in the WS. Namely, states are arranged into three hierarchical tiers: (i) core states (CS), which are politically functional, economically advanced, and reap most of the wealth; (ii) semi-peripheral (SP) states, which are less well off than CS, but still functional; and (iii) peripheral areas (PA), where inexistent indigenous states (e.g., colonies, failed states) or limited autonomy (as in neo-colonialism) together with a backward and inefficient economy manage to reap only a tiny fraction of WS's wealth. Crucially, a country's position in this hierarchy is relatively 'sticky', but not fixed. Thus, most of the 16$^{th}$ century's core states remain central today.

Yet, some CS have become SP, and some SP ones fell in the PA (and, somewhat more rarely, the opposite). On the whole, these tiers should be considered as clusters around more or less well-reasoned, arbitrary boundaries along a core-periphery continuum (Hopkins and Wallerstein 1977).

Conceptually, core countries use sanctions to deter states on other tiers from asserting their interests in the WS. Formally, the theory suggests that in a WS perspective:

(A) **Sanctions in a WS perspective**
$\mathcal{H}_{a1}$  sanctions can worsen the target's positions in the world-system

Practically, successful sanctions shift the target's position in the WS over time by worsening its economic performance and, secondarily, inducing political instability.

**4. LITERATURE REVIEW**

Existing models and theories of international sanctions cannot predict with reasonable certainty that when the economic effect materialises and why sometimes it does not happen.

Partly, this is because the most common approach consists in focusing on the sanctioning state/s and the target.

And, in so doing, the literature misses on the *network effects* that sanctions can ignite.

But, as richer economies get sanctioned, the benefit that third countries derive from participating in their endeavours to prepare for, react to, and evade those measures increases.

Moreover, current methods struggle to account contemporaneously for these network effect and other WS-level processes whose relevance for international sanctions' effectiveness this paper hypothesises:

(A) **Sanctions in a WS perspective (continued)**
$\mathcal{H}_{a2}$  sanctions against the RF failed because they induced a shift in trade ties
(B) **Network effects:**
$\mathcal{H}_{b1}$  regional allies and non-aligned countries can derail the sanctions
$\mathcal{H}_{b2}$  decoupling both weakens sanctions and is enhanced by them
(C) **World – system implications:**
$\mathcal{H}_{c}$  sanctions against Russia contributed to the international decoupling

Remarkably, scholars from the 'West' argue that backfilling by potential spoilers is ineffective 'as a political and economic strategy' (Mau 2016, 358) due to the 'fear of US penalties' (Lukin 2021, 336). Yet, a chronic lack of data supporting this stance makes ignoring populous and growing countries like the People's Republic of China (PRC) and India a rather hasty choice (Morgan, Syropoulos, and Yotov 2023, 22–23).

Moreover, most studies ignore that some countries have been sanctioned for years or decades already. Thus, they accumulated invaluable experience in countering the sustained weaponization of economic tools (Smagin 2022). And the sanctions against the RF have been giving a boost to economic decoupling. For instance, fearing a possible ban of Russian banks from foreign credit-card circuits, the Russian Central Bank began working on MIR, an autochthonous payment system, in 2014 (Kochergin and Yangirova 2018). By 2022, MIR went from being a pre-emptive countermeasure to helping Russia circumvent sanctions. Predictably, other sanction-hit economies, as well as non-aligned countries and even some US allies have expressed interest in MIR (cf. Romanova 2022).

## 5. DATA AND METHODS

The adoption of network methodology in IPE and, more precisely, WS analysis began in the last quarter of the 20th century (Snyder and Kick 1979; Nemeth and Smith 1985). The reason for this combination is that network science is 'uniquely equipped' to grasp the political-economic networks of flows between the WS's constituent actors as well the latter's positions in these networks and the pattern of flows between these positions (Smith and White 1992, 858). Namely, networks allow to represent in a coherent and easily accessible way the most important traits of the WS from the point of view of IPE (Snyder and Kick 1979, 1103): (i) discrete tiers along a continuum; (ii) diachronic movement between/within tiers; (iii) patterns of interaction between tiers; (iv) Matthew effect and wealth-gap.

Formally, this is made possible by using networks to represent units (denoted as the set of vertexes $\mathcal{V} = \{v_1, v_2, ..., v_n\}$) and the interactions amongst them (the set of ties $\mathcal{E} = \{e_1, e_2, ..., e_m\}$). To investigate the world-system effect of international sanctions, with a focus on the RF since 2014, this paper deploys data on international trade in goods and service from the *Atlas of economic complexity* (Harvard Growth Lab [2013] 2022). The dataset provides figures for imports and exports between 1962 and 2020 ($\mathcal{T} = [1962, 2020]$) amongst a total of 249 unique countries. Then, three networks were obtained from this data using an innovative network-construction algorithm: exports ($\mathbf{X}$), imports ($\mathbf{I}$), and net exports ($\mathbf{NX}$). First, given that trade flows vary massively in value across countries, the weight of the ties between countries (the set $\mathcal{W} = \{w_1, w_2, ..., w_m\}$) was normalised so that the value of all imports or (net) exports from any contry $v_i$ sums to the unit ($\sum W_{v_i} = 1$). Intuitively, the sum of all weights equals the number of countries participating in international trade in that year ($\sum W_t = n_t, \forall t \in \mathcal{T}$, with $154 \leq n_t \leq 236$).

Then, in order to reduce the 'noise' in the network, all ties weighting less than five percent of the total flow were omitted (($w_k < .05 \leftarrow 0) \Rightarrow W_t \in [.05, 1]$).[2] Each of the resulting networks detail more than 800,000 interactions between more than 12,000 units (249 unique sovereign political entities) over 59 years.

Clearly, the staggering size of the networks under analysis makes a traditional sociometric analysis intractable. So, the paper relies on blockmodeling, a statistical technique for the simplification of large, potentially incoherent networks into smaller, intelligible structures (on blockmodeling see: Snijders and Nowicki 1997). Namely, this paper uses stochastic blockmodeling (SBM), which decomposes the network into groups of units that have similar pattern of ties called *clusters*. Notably, countries grouped in the same cluster are *stochastically equivalent* and, thus, play a similar role in the network. Practically, the analysis was ran using an approach that combines SBM 'with independent Markov chains for the evolution of the nodes groups through time.' (Matias and Miele 2017, 1121)[3] True, this approach has some limitations (cf. Cugmas and Žiberna 2023, 18ff), and requires a constant number of cluster (see Figure 1). However, it captures both tier-switching and decoupling dynamics by allowing units to change group memberships and the connections between clusters to vary over time (Matias and Miele 2017, 1122–24). Consequently, unlike mainstream methods to study sanctions, it is well suited an holistic analysis of sanctions.

## 6. FINDINGS: RUSSIA'S POSITION IN THE WS THE EFFECT OF SANCTIONS

A topographic analysis of the networks shows that the RF's position in the WS has been evolving sine the first sanctions were imposed in 2014, but not in the direction often supposed. Mainly, the RF had more ties representing less than five percent of the total trade flow than the rest of the world. Numerically, the ratio between minor and major flows,[4] is larger than the world average for RF by $18 - 23\%$. Moreover, the concentration indexes of removed and present ties (plus other indicators not shown due to page limits) suggest that these averages mask a diachronic difference. Basically, the RF's trade ties diversified faster and more than the global average since 2015, suggesting an ongoing attempt at decoupling from the core, sanctioning states. Thus this data contradicts the vulgate that the RF was unable to pivot towards other markets while under international sanctions (e.g., Mau 2016; Lukin 2021).

---

[2] This operation could potentially reduce the number of units in the network by cancelling out a country that has more than 20 ties, none of which representing at least five percent of its total trade. However, this case did not materialise in this dataset.

[3] This SBM is implemented in the add-on package dynsbm for the statistical programming language R.

[4] Calculated as $(m - \delta)/m$, with $\delta = \text{count}(w_j < .05)$.

**Figure 1.** Results of the dynamic SBM on all networks (2014–2020)

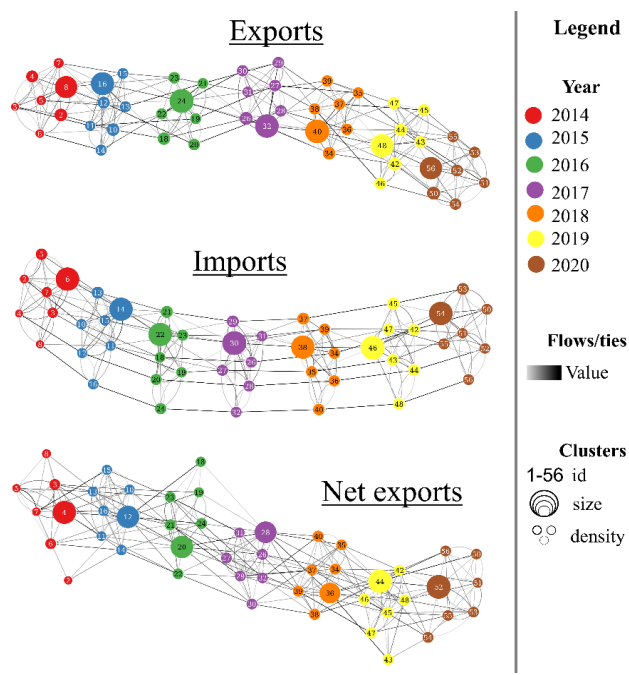

*Source:* Author's elaboration using the Python module graph-tool and Paint.net.

Moving to the blockmodeling results, the dynamic SBM shows verifies all hypothesis by showing that (1) sanctions may worsen the targets' position and even downgrade them to a lower tier ($\mathcal{H}_{a1}$); (2) but this does not always happen. Also, (3) the RF's position has not changed ($\mathcal{H}_{a2}$), albeit its ties to the 'Western' CS ($\mathcal{H}_{b1}$) and the SP ($\mathcal{H}_{b2}$) morphed in the last decade inducing significant re-arrangements in the lower tier. Overall, (4) these shifts seem to constitute the incremental steps of an ongoing global decoupling within the WS ($\mathcal{H}_c$) that sees the PRC, the RF and other sanction-hit states form a parallel core and SP.

Yet, the presence of seven clusters across all time periods, for a total of 49,[5] rather than three warrants a preliminary explanation. Essentially, the WS's tiers are a function of guesstimates regarding the exact demarking of core and (semi-)periphery along a continuum (see above; and Hopkins and Wallerstein 1977). Thus, there is an internal diversification between the tiers into lower/upper subsets of states. Moreover, the three tiers may not be geographically contiguous (Wallerstein 1976, 230). But proximity does influence international trade (cf. Anderson 2010). Hence, states tend to trade more with neighbouring countries regardless of their tier. So, more than three clusters are needed to account for economic stratification within the tiers as well as geographical agglomeration across them.

### 6.1. Sanctions may affect targets' position in the WS…

Before focusing on the case of the RF, it is worth highlighting that the hypothesis that successful sanctions affect the target's position in the WS is verified by the data. For instance, most sanctions against Iran were lifted after the signing of the Joint Comprehensive Plan of Action (JCPOA). Thus, in 2012. in 2015–18, Iran came out of international isolation and sat in the mid-to-low SP cluster (id: 2, 10, 18, 26, 34) along Vietnam, Ukraine, and Serbia. However, in 2018–19, the US's withdrawal from the JCPOA marked the start of the Trump administration's 'maximum pressure' campaign (see Brewer and Nephew 2019). As a result, Iran's position in the WS worsened, with Teheran dropping into the PA since 2019 (clusters: 48, 56).

Similarly, the US-Cuba thaw during the Obama administration allowed Havana to sit in a SP cluster (id: 4, 12; spanning Latin-America and the Caribbeans) in 2014-2015. But the US's tougher stance under Trump led to tighter sanctions and Cuba's drop into the PA (clusters: 24, 43, 40, 48, 46).

### 6.2. … but it does not always happen

Still, not all sanctions seem to have the same effect and some targeted countries keep their position.

---
[5] Note: The clusers in Figure 1 are labelled 1-56 and the the first cluster of each time period (1, 9, 17, etc.) is homitted because some countries in the dataser do not exist anymore (e.g., Yugoslavia).

Indeed, this could be the case due to the prevalence of regional trade or a key role in markets with tight supply. The network effects due to productive complementarities amongst sanctioned states and their non-sanctioning neighbours are evident in the case of Venezuela. The country is under Western sanctions since 2017 (Buxton 2018). However, its position in the WS has remained stable as it most of its trade in sanctioned goods diverted towards non-hostile countries. So, Caracas resisted the sanctions thanks to trading scheme with neighbouring neutral states and deepening ties with India and the PRC. On a larger scale and over a longer period, the same processes applied to the RF. Practically, it remains in the cluster of European CS where it was in 2014, with France, Germany, Italy, and the Netherlands (id: 6, 14, 22, …, 54). True, Moscow's global standing makes the explanation for sanctions' ineffectiveness much more complex than in Venezuela's case (see para. 6.3, below). However, both factors characterising the former case are also present here. So, the RF benefitted from continued trade with regional allies within the Eurasian Economic Union (EEU).[6] Moreover, the data shows that trade shifted from the EU core and SP states (id: 2, 10 18, …, 50) towards friendlier countries, marking an incremental decoupling. Yet, the RF's trade is much more intensely global than Venezuela's, extending beyond regional allies in the EEU to friendly BRICS countries and non-hostile ones like Egypt, Turkey, and Qatar.

### 6.3 The RF took advantage of re-arrangements in the semi-periphery

Intuitively, the effect of sanctions against the RF is much more relevant to the WS due to its position in the core. Essentially, the RF's pivot towards new trade partners, some of which are also geopolitical allies, brought about visible changes in the WS. Arguably, international sanctions catalysed the long-term, structural processes leading to these shifts that originated with the US's decline and the PRC's rise the only *primus inter pares* along the US (clusters: 3, 11, 19, …, 51). Basically, countries tightly connected with the RF rose in the ranks of the WS as the former diverted its trade away from sanction senders. Most notably, as its cooperation with the RF intensified, Turkey has moved the mid-low SP to the cluster of European CS in 2018. Converesely, the RF's behaviour penalised Saudi Arabia, a SP commodity exporter. Arguably, the RF's discounts on hydrocarbons to friendly countries contributed to undercutting the Saudis.

However, mere dumping from a much smaller oil producer cannot explain Riyad's fall of grace from the Asia-Pacific cluster in 2014–16 (id: 5, 13, 21) down to the lower SP (cluster: 44, 52) in 2019–20. Rather, it seems that the RF's decoupling due to international sanctions spoiled the Saudis' somewhat awkward attempt to dance with two partners in its relations with the US, the PRC and the RF. Coherently, the data shows that, as sanctions against the RF toughened, the PRC substituted a reliable Moscow for an ambivalent Riyad. Meanwhile, the US and its allies weaponised efforts to reduce fossil-fuel use to show their dissatisfaction with Saudi Arabia's winking at strategic competitors.

### 6.4 From decoupling to the emergence of a parallel core and semi-periphery

Overall, the SBM shows that international sanctions on a CS can lead to more or less sudden decoupling with visible effects on SP states and the PA. Besides those mentioned above, it is worth mentioning also Egypt's temporary drop-out from the mid-to-low SP cluster into the PA in 2019 before relocating in the Latin-America SP in 2020. Interestingly, this period saw Cairo cooperating closely with the RF and distancing itself from the US somewhat. Still, many movements within SP clusters and to the PA are not directly connected to sanctions against the RF: e.g., Hong Kong following Egypt's erratic trajectory in the same years. Thus, it is necessary to contextualise the shifts associable with the reorientation of the RF's trade within the wider transformation of the WS due to the PRC's rise. After all, the PRC is the only other 'basketball ball' whose pattern of ties can dramatically change the global structure of the entire trade network. In summary, the SBM suggests that the Latin-American periphery is slowly turning into a trans-continental cluster of non-aligned SP states. Symmetrically, the mid-to-low SP cluster gather SP countries aligned or allied with the US. Meanwhile, the dualism within the *primus inter pares* cluster is mirrored in the European-CS cluster. Interestingly, models with more clusters separate this group into two distinct cores separating US-allied European CS (France, Germany, Italy, the UK, and the Netherlands) from the RF and an ambivalent Turkey. Meanwhile, SBMs with less clusters merge these clusters along geopolitical cleavages, separating the US and its allies from the PRC, the RF, and Turkey. Summarily, these results show that a parallel core-SP structure is emerging in the WS. And international sanctions against the RF are accelerating a previously slow process of incremental decoupling between US-allied and non-aligned CS and SP states.

---

[6] Besides Russia, the EEU includes Armenia, Belarus, Kazakhstan, Kirghizstan, and Uzbekistan.

## CONCLUSION

This paper provides the first systematic account of international sanctions in a world-system perspective focusing on the case of the sanctions against the Russian Federation. Namely, it analyses the network of global trade (2014–20) with state-of-the-art stochastic blockmodeling (SBM) for dynamic networks (Matias and Miele 2017) to capture a series of *network effects* connected these sanctions. Practically, this dynamic SBM is applied on a network of exports built on high-quality data from the *Atlas of Economic complexity* (Harvard Growth Lab [2013] 2022) using an innovative network-creation algorithm. Arguably, the dynamics that detailed the sanctions against Russia can be identified also in other cases (e.g., those against Venezuela): trade with regional allies and shift to non-hostile trade partners. However, theory-building and generalisations should be left to further research. Also, subsequent researches may remedy this paper's limitation in several way. First, by using data covering the sanctions connected with the Russo-Ukrainian war and the Biden administration's sanctions against the PRC. Second, given that the dynamic SBM used for this paper does not allow for changes in the number of clusters over time, by splitting the dataset's timespan in theoretically sound time periods associated with different numbers of clusters. Third, by exploring models with more/less clusters to shed a light on the evolution of specific sub-systems (e.g., the Asia-Pacific region) or highlight differences and similarities between several block models (potentially obtained with different SBM approaches). More limitedly, this paper asks how did the sanctions against Russia affect the world-system, if at all. On the basis of a thorough literature review, this question is operationalised by formulating three sets of hypotheses: (a) sanctions can worsen the target's positions in the world-system, but (a-2) sanctions against Russia did not achieve this goal; rather they induced a re-arrangement of Russia's trade flow (b-1) away from sanction senders (b-2) towards non-hostile countries; and (c) sanctions against Russia contribute to the decoupling of non-aligned economies form the US-led core and SP states. Eventually, the SBM and a topographical analysis reveal that all three sets of hypotheses are verified. Most importantly, sanctions against Russia both failed because of and contributed to enhance network effects amongst (actually and potentially) sanctioned states. Overall, sanctions and the wealth-shift to Asia foreshadow a parallel core-SP structure that could sidestep US hegemony.

**SUMMARY**

The existing literature on international economic sanctions has not addressed the issue from the perspective of world-system analysis. Yet, dependency/world-system theory is uniquely apt to the investigation of political-economic ties amongst sovereign political units. Moreover, this approach's relational ontology combines perfectly with network-analysis methodology. Indeed, networks have already been used to study the world-system in the last quarter of the 20[th] century, but not in relation to sanctions specifically. Furthermore, both fields have advanced significantly since then making previously intractable statistical analysis possible and providing words for previously unknown geoeconomic phenomena. Against this background, this paper contributes to fill this gap by using analysing the network of global trade (2014–20) with state-of-the-art stochastic blockmodeling (SBM) from dynamic networks to capture a series of *network effects* connected with international sanctions against the Russian Federation. Practically, this dynamic SBM is applied on a network of exports built on high-quality data from the *Atlas of Economic complexity* using an innovative network-creation algorithm. The paper asks how, if at all, did the sanctions against Russia affect the world-system. On the basis of a thorough literature review, this question is operationalised by formulating three sets of hypotheses: (a) sanctions can worsen the target's positions in the world-system, but (a-2) sanctions against Russia did not achieve this goal; rather they induced a re-arrangement of Russia's trade flow (b-1) away from sanction senders (b-2) towards non-hostile countries; and (c) sanctions against Russia contribute to the decoupling of non-aligned economies form the US-led core and SP states. Eventually, the SBM and a topographical analysis reveal that all three sets of hypotheses are verified. Most importantly, sanctions against Russia both failed because of and contributed to enhance network effects amongst (actually and potentially) sanctioned states. Overall, sanctions and the wealth-shift to Asia foreshadow a parallel core-SP structure that could sidestep US hegemony.